\documentclass[preprintnumbers,amsmath,amssymb,nofootinbib]{revtex4}
\usepackage[T1]{fontenc}
\usepackage{amssymb}
\usepackage{latexsym}
\usepackage{color}
\usepackage{graphicx}
\usepackage{hyperref}
\usepackage{color}

\def\nn{\nonumber\\}
\newcommand{\f}[2]{\frac{#1}{#2}}
\def\be{\begin{equation}}
\def\ee{\end{equation}}
\def\bea{\begin{eqnarray}}
\def\eea{\end{eqnarray}}

\begin{document}
\title{Gravitational Collapse of an Inhomogeneous Fluid in Rastall Theory}
\author{Akbar Jahan\footnote{jahan@riaam.ac.ir}, Naser Sadeghnezhad\footnote{nsadegh@riaam.ac.ir}, Amir Hadi Ziaie\footnote{ah.ziaie@riaam.ac.ir}}
\address{Research~Institute~for~Astronomy~and~Astrophysics~of~ Maragha~(RIAAM), University of Maragheh,  P.~O.~Box~55136-553,~Maragheh, Iran}
\begin{abstract}
We study spherically symmetric gravitational collapse of an inhomogeneous fluid with anisotropic energy momentum tensor (EMT) in Rastall gravity. Considering a linear equation of state (EoS) for the fluid profiles, i.e., $p_r=w_r\rho$ and $p_\theta=w_\theta\rho$, we try to build and investigate non-singular collapse scenarios for which, the spacetime singularity that appears in the homogeneous case~\cite{ahz2019}, is absent. We therefore set the Rastall parameter in such a way that the effective radial pressure vanishes. This helps us to obtain a class of exact nonsingular solutions in which the matter shells undergo a {collapse} process in a contracting regime, reach a bounce point, and then enter an expanding phase. We further investigate formation of trapped surfaces during the dynamical evolution of the collapsing body. {It is found} that for the obtained solutions, {the} trapped surface formation can be avoided and consequently, the bounce event is not covered by the apparent horizon. Validity of weak energy condition (WEC) is also examined for the obtained solutions.
\end{abstract}

\maketitle
\section{Introduction}
The final fate of gravitational collapse of a dense body under its own gravity is one of the most mysterious and challenging problems in gravitation and astrophysics. Basically, the gravitational collapse of a stellar body occurs when the inward pull of gravity overcomes the outward forces like the thermal pressure {of} nuclear fusion or {the degeneracy pressure} of electrons and neutrons. Several scenarios can lead to {the} gravitational collapse, depending on the mass and evolutionary stage of the star: $i)$ exhaustion of nuclear fuel (for main-sequence and post-main-sequence stars)\cite{exhnucfu}, $ii)$ overcoming electron (Chandrasekhar limit) or neutron (Tolman-Oppenheimer-Volkoff limit) degeneracy pressures~\cite{elneudeg}, $iii)$ external perturbations such as mass accretion from a companion star or binary mergers~\cite{massacrbin}\cite{massacrbin1} and $iv)$ general instability such as pair instability\footnote{In extremely massive stars (100$M_\odot$), photon pressure can produce electron-positron pairs, reducing thermal pressure and triggering collapse~\cite{pairinst}.} and hypothetical dark matter accumulation~\cite{darkmatter}. See also~\cite{handsuper} for more details. In the context of general theory of relativity (GR), much of what we have learned about the endstate of a gravitational collapse process is due to the celebrated singularity theorems of Geroch, Hawking and Penrose~\cite{HAWPENST}\cite{CCCREF}. These theorems predict that the collapse gives
rise to the formation of a spacetime singularity, a spacetime event at which the densities and spacetime curvatures
get arbitrarily large values and diverge {due to ultra-strong gravitational field}. A singular region of the spacetime can either be hidden within an event horizon of gravity (black hole formation) or visible to the external Universe. In the former, the horizon forms earlier than the singularity formation and in the latter the horizon is delayed thus allowing null geodesics to escape the singularity and reach faraway observers (naked singularity formation)~\cite{Joshibook}\cite{ColReview}. The occurrence of a singular event within the spacetime is associated with crucial problems for the issues such as physical determinism, predictability, path incompleteness and validity of the standard laws of physics~\cite{singularpr}. The first model of gravitational collapse was proposed almost eighty years ago by Oppenheimer, Snyder and Datt (OSD)~\cite{OSDM}, where as an attempt to model an idealized stellar object, they studied the dynamical collapse of a homogeneous spherical dust cloud under its own weight. In the OSD model, the collapse process culminates in the formation of a spacetime singularity necessarily dressed by an event horizon, as hypothesized by the cosmic censorship conjecture (CCC)~\cite{CCCREF}. Despite its simplicity and ideality, the OSD model paved the way for studying the collapse process in the presence of more general conditions such as the inclusion of inhomogeneities within the density and pressures of the collapsing cloud. In analyzing \textcolor{black}{the} gravitational collapse, inhomogeneous models play a crucial role as they describe scenarios where the collapsing matter exhibits non-uniform distributions. These models offer a more accurate framework for studying \textcolor{black}{the} collapse dynamics, since they incorporate both temporal evolution and spatial variations in \textcolor{black}{the} density and pressures~\cite{inhommods}. Such collapse processes frequently result in shell-focusing singularities which are true spacetime singularities where \textcolor{black}{the} matter shells collapse to zero radius, with the singular epoch varying radially due to \textcolor{black}{the} inhomogeneities~\cite{Joshibook}. Extensive research over the past years has revealed that, under specific conditions, these singularities may become visible, thereby challenging the CCC~\cite{CCCREF}\cite{CCCREF1} and raising fundamental questions about the predictability and determinism of GR~\cite{inhommods11}\cite{inhommods2}\cite{COUNTERCCC}, see also~\cite{ColReview}\cite{trapsurf} and references therein.
\par
\textcolor{black}{The} spacetime singularities are commonly known to be problematic since the classical framework of GR fails to be predictive and breaks down when \textcolor{black}{the} basic quantities become infinite. In this sense, the occurrence of a spacetime singularity implies that the theory has been applied beyond its domain of validity~\cite{GRSINDOM}\cite{GRSINDOM1}. On one hand, it is generally believed that at \textcolor{black}{the} singular regions of spacetime where \textcolor{black}{the} super-dense regimes of extreme gravity are present, a quantum theory of gravity would be the most likely description of the phenomena created by such spacetime events~\cite{trapsurf}\cite{Moore2006}. Hence, in quantum-corrected collapse models, the effects of quantum gravity modify the classical dynamics to prevent singularity formation leading to a bounce or a smooth non-singular endstate~\cite{singremove}\cite{quantumcollapse}, instead. On the other hand, many efforts have been made during the last decades to cure the singularity problem by considering alternative theories of gravity. The results of research in this arena show that in these theories, the contribution due to additional/correction terms can help avoiding singularities that are present in the framework of GR. For example a nonsingular model of the collapse process in $F(R)$  gravity is given in~\cite{FRsingbamba} where the authors found that the addition of $R^\alpha (1<\alpha\leq2)$ term could cure the curvature singularity leading thus to a model free of \textcolor{black}{such a problem}. Also, the authors of~\cite{modbd2018} have examined the possibility of singularity avoidance in \textcolor{black}{a} generalized Brans-Dicke (BD) theory with a running coupling parameter. In the context of Einstein-Cartan theory it has been shown that the effects of a spin source can avoid the curvature singularity, \textcolor{black}{replacing} it by a non-singular bounce~\cite{ecnonsing}. \textcolor{black}{In}~\cite{bambi2014} it is argued that \textcolor{black}{the} classical collapse can be affected by \textcolor{black}{a} non-minimal coupling between \textcolor{black}{the} gravity and fermions long before the effects of quantum gravity become dominant. Very recently, the effects of non-vanishing spacetime torsion on the final fate of the collapse scenario in BD theory has been investigated in~\cite{BDtorsion2024}. Work along this line has been also extended to \textcolor{black}{the} cosmological scenarios where non-singular bouncing models have been presented in modified gravities~\cite{modgrnonsing}. 
\par
Among the modified gravity theories, Rastall theory has also benefited from a careful study of homogeneous collapse process~\cite{ahz2019}\cite{highdrascol}. In contrast to GR, in Rastall theory, the ordinary energy-momentum conservation law ($\nabla_{a}{\rm T}^{ab}=0$) does not hold anymore~\cite{Ras1972}. Such a modification to GR which has received phenomenological confirmations by particle production process in cosmology~\cite{partcreat} was firstly proposed by Peter Rastall~\cite{Ras1972}. The basic motivation for such \textcolor{black}{an} assumption was that  \textcolor{black}{the} conservation laws have been only tested on the Minkowski background spacetime or quasistatic gravitational fields~\cite{Ras1972}\cite{minquasifields} and hence, they may no longer hold in  \textcolor{black}{a} curved spacetime. According to Rastall's assumption, the usual conservation law of the EMT is modified as $\nabla_{a}{\rm T}^{ab}=\lambda\nabla^{b}{\mathcal R}$, where ${\mathcal R}$ is the Ricci curvature scalar and $\lambda$ is a constant which represents the non-minimal coupling between the geometry and \textcolor{black}{the} energy-momentum tensor~\cite{coupsource}. However, during the past years, Rastall gravity has been criticized by some researchers \textcolor{black}{arguing} that \textcolor{black}{the} Rastall theory is fundamentally equivalent to GR~\cite{Visser2018}\cite{Golovnev2024}. The arguments and supports that have been put forward in favor of this theory can be summarized as follows: $i)$ While the field equation of Rastall gravity may be mathematically recast into GR-like form, the physical meaning of the EMT in this theory is different, see e.g.,~\cite{Rasfart} for similarities (through the Lagrangian formulation) between \textcolor{black}{the} Rastall gravity and theories with curvature-matter coupling such as $f(R,T)$ and $f(R,{\mathcal L}_m)$ theories. Indeed, the non-conservation of EMT in \textcolor{black}{the} Rastall theory leads to non-trivial interactions between \textcolor{black}{the} geometry and matter fields, which are not present in GR~\cite{morad2018}. Physically, this means that \textcolor{black}{the} matter fields can exchange energy with the gravitational field itself, and consequently, such a mutual interaction can result in novel interesting implications in cosmological or astrophysical models~\cite{Rascosmols}\cite{Rascosmols1}\cite{validras}\cite{validras1}. $ii)$ \textcolor{black}{the} Rastall gravity explicitly violates the standard conservation law which is the cornerstone of GR. Such \textcolor{black}{a} violation has implications for black hole physics, gravitational lensing effects, black hole shadows and quantum field theory in curved spacetime~\cite{Ras1972}\cite{partcreat}\cite{minquasifields}\cite{coupsource}\cite{validras}\cite{glensras}\cite{zhung2021}. See also~\cite{morad2018} and references therein. $iii)$ \textcolor{black}{the} Rastall theory can deviate in cosmological scenarios with perfect fluid as the matter content, for example, the equations of state of dark matter or dark energy may differ in \textcolor{black}{the} Rastall framework~\cite{chagua2023}. Also, stellar structure models such as neutron and quark stars could exhibit deviations due to the modified conservation law~\cite{stellras}. $iv)$ \textcolor{black}{the} Rastall gravity offers a simpler and more tractable alternative in comparison to other modified gravity theories while still introducing significant deviations from GR. Whereas many modified theories such as Brans-Dicke, Einstein-Gauss-Bonnet or Horndeski theories introduce complicated higher-order field equations or require additional scalar fields, in \textcolor{black}{the} Rastall framework the non-conservation term provides a single adjustable parameter to govern deviations from standard GR~\cite{BDtorsion2024}\cite{modgrtheos}. This makes it particularly useful for obtaining \textcolor{black}{the} exact analytical solutions in scenarios like gravitational collapse, where \textcolor{black}{the} other theories often necessitate numerical approximations.
\par 
Recently, in~\cite{nonsinrasgen} the collapse of a homogeneous perfect fluid has been studied in a generalized version of Rastall theory~\cite{genras}. The authors of~\cite{nonsinrasgen} have assumed a varying coupling parameter, $\lambda\rightarrow\lambda(t)$, and showed that the singular OSD model is modified to a nonsingular bouncing scenario. Motivated by this work, we here consider the \textcolor{black}{gravitational} collapse of an anisotropic inhomogeneous fluid in Rastall gravity and examine whether nonsingular solutions can exist. The organization of present work is as follows: In Sec.~\ref{sec1}, we give a brief review on Rastall theory and find the corresponding field equations that govern the collapse process. Sec.~\ref{sec2}, deals with the study of exact nonsingular solutions and Sec.~\ref{sec3}, is devoted to investigating the trapped surfaces. Finally, Sec.~\ref{sec4} is about .the conclusions.
\section{Field equations of Rastall gravity}\label{sec1}
To investigate the collapse process in \textcolor{black}{the} Rastall gravity we follow the Rastall's proposal, that is, the covariant divergence of the EMT is no longer zero but is proportional to the covariant derivative of the Ricci curvature. Such a generalization on EMT conservation \textcolor{black}{law}, leads to the modification of Einstein tensor as~\cite{Ras1972}\cite{Rascosmols}
\be\label{Einsmod}
G_{ab} \rightarrow G_{ab}+\kappa_{r}\lambda g_{ab}{\mathcal R},
\ee
where $\kappa_{r}$ is a constant. The Rastall field equation then reads
\be\label{RastallFES}
{\rm G}_{ab} +\gamma g_{ab}{\mathcal R}=\kappa {\rm T}_{ab},
\ee
where $\gamma=\kappa_r\lambda$ is the Rastall parameter. The above equation can be rewritten in an equivalent form with an effective source as
\be\label{FESEquiv}
{\rm G}_{ab}=\kappa_{r}{\rm T}^{\rm eff}_{ab},~~~~{\rm T}^{\rm eff}_{ab}={\rm T}_{ab}-\f{\gamma g_{ab}{\rm T}}{4\gamma-1}.
\ee 
The static weak field limit gives the following relation 
\be\label{stst} 
\kappa_r\f{6\gamma-1}{4\gamma-1}=\kappa,
\ee 
where $\kappa=8\pi G/c^2$ is the Einstein's gravitational constant. In the weak field limit with $\lambda=0$, we have $\kappa_r=\kappa$ therefore the standard Einstein gravity is recovered. This can also be deduced from Eq.~(\ref{FESEquiv}) where we observe that the Rastall parameter acts as a measure of \textcolor{black}{the} non-minimal coupling between matter (trace of EMT) and geometry (spacetime metric). Such a coupling is usually understood as mutual interaction between matter and geometry. It is therefore our aim here to build and study nontrivial collapse models in the presence of such an interaction. We then examine \textcolor{black}{the} gravitational collapse of type (I) matter fields~\cite{HAWPENST} for spherically symmetric spacetimes considered here. This form of matter along with its corresponding EMT is utilized to describe most of the physically reasonable matter fields (including dust, perfect fluids, massless scalar fields and such others, see e.g.~\cite{maeda2021} and the discussions therein), except the special cases that are described by type (II) matter fields. The EMT of a type (I) matter can be written as
\be\label{Tmunu}
{\rm T}^{ab}=\lambda_1E_1^aE_1^b+\lambda_2E_2^aE_2^b+\lambda_3E_3^aE_3^b+\lambda_4E_4^aE_4^b,
\ee
where $(E_1,E_2,E_3,E_4)$ is an orthonormal basis with $E_4$ being timelike eigenvector and $\lambda_i(i=1,2,3,4)$ are the eigenvalues. For a spherically symmetric matter distribution, the spacetime geometry that describes the collapsing cloud is given by the general metric in the comoving coordinates $x^i=\{t,r,\theta,\phi\}$, as
\be\label{met}
ds^2=-{\rm e}^{2\nu(t,r)}dt^2+{\rm e}^{2\psi(t,r)}dr^2+R^2(t,r)d\Omega^2,
\ee
where $d\Omega^2$ is the standard line element on the unit two-sphere and $\nu$, $\psi$ and $R$ are functions of $t$ and $r$. Here $R(t,r)$ is the physical area radius and
$r$ is the shell labeling comoving radial coordinate. Hence, the quantity $4\pi R^2(t,r)$ is equivalent to the proper area of the mass shells and the area of such a shell with $r=cte.$ vanishes when $R(t,r)\rightarrow0$. Therefore, the curve $R(t,r)=0$ describes the spacetime singularity, an event at which all the mass shells with $0\leq r<\infty$ collapse to a vanishing volume. This type of singularity is usually known as a shell-focusing singularity~\cite{Joshibook}. The EMT which is given by (\ref{Tmunu}) admits only diagonal components in the comoving coordinate system. The non-vanishing components of the EMT are then given by
\bea\label{emtcomps} 
{\rm T}^{t}_{\,t}&=&-\rho(t,r),~~~{\rm T}^{r}_{\,r}=p_r(t,r),~~~\nn {\rm T}^{\theta}_{\,\theta}&=&p_\theta(t,r)={\rm T}^{\phi}_{\,\phi}=p_\phi(t,r),~~~{\rm T}^{t}_{\,r}={\rm T}^{r}_{\,t}=0,
\eea
where $\rho$, $p_r$, $p_\theta$ and $p_\phi$ are the eigenvalues of ${\rm T}^a_{\,b}$ and are interpreted as the density and principal pressures. In realistic stellar models the pressure profiles are expected to be anisotropic. This pressure anisotropy may arise owing to several factors such as strong magnetic fields, rotational forces, superfluidity or phase transitions in the stellar material~\cite{pressaniso}. See also~\cite{das2025} for recent reviews. We therefore \textcolor{black}{consider} anisotropic pressure profiles to have a collapse model, as close as possible to realistic models\footnote{In fact, the simplest reason for the presence of anisotropy in a self-gravitating object is the shearing motion of the fluid~\cite{shearjoshi}. As the spacetime metric~(\ref{met}) admits non-vanishing shear tensor in the form
\be\label{shearten}		
\sigma^{\theta}_{\,\theta}=\sigma^{\phi}_{\,\phi}=-\f{1}{2}\sigma^{r}_{\,r}=\f{1}{3}{\rm e}^{-\nu}\left[\dot{\psi}-\f{\dot{R}}{R}\right],
\ee		
it is then expected that due to the spacetime shear, the pressure profiles are different in radial and tangential directions, see also~\cite{shearrefs} and references therein}.
The matter fields are usually taken to satisfy the WEC, i.e., the energy density as measured by any observer in spacetime is non-negative. Then, for an observer with four-velocity $v^\alpha$ the inequality ${\rm T}_{ab}v^av^b\geq0$ must hold where, $v^a$ is any future-directed timelike vector field. For a type (I) matter the WEC is fulfilled, provided that~\cite{HAWPENST}
\be\label{WEC}
\rho\geq0,~~~~~~~\rho+p_j\geq0,~~~j=\{r,\theta,\phi\}.
\ee
Furthermore, another energy condition is believed to hold for physically reasonable matter fields. This condition which is known as the dominant energy condition (DEC) states that \textcolor{black}{the} matter distribution should follow timelike or null world lines~\cite{toolkitpoisson}. If $u^a$ is any timelike future-directed vector field, then the DEC requires that the inequality ${\rm T}_{ab}u^au^b\geq0$ holds and ${\rm T}_{ab}u^a$ is non-spacelike. Also, the strong energy condition (SEC) asserts that for any timelike future directed vector field $w^a$ the inequality ${\rm T}_{ab}w^aw^b\geq-{\rm T}/2$ must be \textcolor{black}{considered}~\cite{toolkitpoisson}. In terms of the EMT components, these two conditions take the following form
\bea
\!\!\!\!\!\!\!\!\!\!\!\!&&{\rm DEC:}~~\rho\geq0,~~~{\rm and}~~~\rho\geq |p_i|,~~~i=\{r,\theta,\phi\},\label{decsec1}\\
\!\!\!\!\!\!\!\!\!\!\!\!&&{\rm SEC:}~~\rho+p_i\geq0,~{\rm and}~\rho+\sum_{i}p_i\geq0.\label{decsec2}
\eea
The components of effective EMT are found as
\bea
{\rm T}^{t\,{\rm eff}}_{\,\,t}\equiv-\rho^{\rm eff}&=&-\f{(3\gamma-1)\rho+\gamma(p_r+2p_t)}{4\gamma-1},\label{EFFEMT1}\\
{\rm T}^{r\,{\rm eff}}_{\,\,r}\equiv p_r^{\rm eff}&=&\f{(3\gamma-1)p_r+\gamma(\rho-2p_t)}{4\gamma-1},\label{EFFEMT2}\\
{\rm T}^{\theta\,{\rm eff}}_{\,\,\theta}={\rm T}^{\phi\,{\rm eff}}_{\,\,\phi}&\equiv& p_\theta^{\rm eff}=\f{(2\gamma-1)p_t+\gamma(\rho-p_r)}{4\gamma-1},\label{EFFEMT3}\\
{\rm T}^{r\,{\rm eff}}_{\,\,t}&=&{\rm T}^{t\,{\rm eff}}_{\,\,r}=0.\label{EFFEMT4}
\eea
We further require that the components of the effective EMT \textcolor{black}{to obey} the WEC\textcolor{black}{, i.e.}
\be\label{WECeff}
\rho^{\rm eff}\geq0,~~~~~~~\rho^{\rm eff}+p^{\rm eff}_j\geq0,~~~j=\{r,\theta,\phi\}.
\ee
Also, the DEC and SEC for effective EMT are given by
\bea
\!\!\!\!\!\!\!\!\!\!\!\!&&{\rm DEC:}~~\rho^{\rm eff}\geq0,~~{\rm and}~~\rho^{\rm eff}\geq |p_i^{\rm eff}|,~~i=\{r,\theta,\phi\},\label{decseceff1}\\
\!\!\!\!\!\!\!\!\!\!\!\!&&{\rm SEC:}~~\rho^{\rm eff}+p_i^{\rm eff}\geq0,~{\rm and}~\rho^{\rm eff}+\sum_{i}p_i^{\rm eff}\geq0.\label{decseceff2}
\eea
For the spacetime metric (\ref{met}) the field equation (\ref{FESEquiv}) along with \textcolor{black}{the} Bianchi identity $\nabla_a{\rm T}^{a\,{\rm eff}}_b=0$ lead to the following set of differential equations~\cite{Joshibook}
\bea
\!\!\!\!\!\!\!\!&{\rm G}&\!\!\!\!^t_t=\kappa_r{\rm T}^{t\,{\rm eff}}_t\Rightarrow\f{\partial}{\partial r}\left[R\left(1-S+H\right)\right]=R^2R^\prime\kappa_r\rho^{\rm eff},\label{fesbia0}\\
\!\!\!\!\!\!\!\!&{\rm G}&\!\!\!\!^r_r=\kappa_r{\rm T}^{r\,{\rm eff}}_r\Rightarrow\f{\partial}{\partial t}\left[R\left(1-S+H\right)\right]=-R^2\dot{R}\kappa_rp_r^{\rm eff},\label{fesbia1}\\
\!\!\!\!\!\!\!\!&{\rm G}&\!\!\!\!^\theta_\theta={\rm G}^\phi_\phi=\kappa_r{\rm T}^{\theta\,{\rm eff}}_\theta\Rightarrow-2\f{\partial^2}{\partial r\partial t}\left[R\left(1-S+H\right)\right]\nn+\!\!\!\!\!\!\!\!&&\!\!\!\!\f{\partial}{\partial r}\left[R\left(1-S+H\right)\right]\f{\dot{S}}{S}+\f{\partial}{\partial t}\left[R\left(1-S+H\right)\right]\f{H^\prime}{H}=4R\dot{R}R^\prime\kappa_rp_\theta^{\rm eff},\label{fesbia2}\\
\!\!\!\!\!\!&{\rm G}&\!\!\!\!^t_r=\kappa_r{\rm T}^{t\,{\rm eff}}_r=0\Rightarrow\dot{R}^\prime-\dot{\psi}R^\prime-\nu^\prime\dot{R}=0,\label{fesbia3}\\
&\nabla_a&\!\!\!\!{\rm T}^{a\,{\rm eff}}_{\,t}=\!\!0\Rightarrow\dot{\rho}^{\rm eff}+\!\left[\dot{\psi}+\f{2\dot{R}}{R}\right]\!\rho^{\rm eff}\!+\!\dot{\psi}p_r^{\rm eff}=-2p^{\rm eff}_\theta\f{\dot{R}}{R},\label{fesbia4}\\
&\nabla_a&\!\!\!\!{\rm T}^{a\,{\rm eff}}_{\,r}=\!\!0\Rightarrow p_r^{\prime\,\rm eff}+\!\left[{\nu}^\prime+\f{2{R}^\prime}{R}\right]\!p_r^{\rm eff}\!+\!{\nu}^\prime \rho^{\rm eff}=2p^{\rm eff}_\theta\f{{R}^\prime}{R},\label{fesbia5}
\eea
where 
\be\label{GH}
S=S(t,r)={\rm e}^{-2\psi}R^{\prime2},~~~~H=H(t,r)={\rm e}^{-2\nu}\dot{R}^{2},
\ee
and $\prime\equiv\partial/\partial r$, $\dot{}\equiv\partial/\partial t$. Eliminating $p_\theta^{\rm eff}$ from Eqs. (\ref{fesbia4}) and (\ref{fesbia5}) along with using Eq.~(\ref{fesbia3}) we get
\be\label{massF}
\f{\partial}{\partial r}\left(p_r^{\rm eff}R^2\dot{R}\right)+\f{\partial}{\partial t}\left(\rho^{\rm eff}R^2R^\prime\right)=0.
\ee
From the above equation, we find that Eqs.~(\ref{fesbia0}) and (\ref{fesbia1}) can be recast into
\be\label{eqsfmass}
\f{F^\prime}{R^2R^\prime}=\kappa_r\rho^{\rm eff},~~~~~~~~~\f{\dot{F}}{R^2\dot{R}}=-\kappa_rp_r^{\rm eff},
\ee
where
\be\label{massdef}
F=F(t,r)=\left(1-S+H\right)R,
\ee
is defined as the mass function. Indeed $F(t,r)$ is the Misner-Sharp mass of the system~\cite{MSE} and is physically interpreted as the amount of matter enclosed in the comoving shell labeled by $r$ at time $t$. In order to preserve the regularity at the initial epoch, we require that $F(t_i,0)=0$, i.e., the mass function should vanish at the center of the collapsing cloud~\cite{Joshibook}. Now, Eq.~(\ref{fesbia4}) together with Eqs.~(\ref{eqsfmass}) give rise to the following result
\be\label{eqtangential}
4RR^\prime\dot{R}{\rm G}^\theta_\theta=-2\dot{F}^\prime+F^\prime\f{\dot{S}}{S}+\dot{F}\f{H^\prime}{H},
\ee
where Eq.~(\ref{fesbia3}) and its derivatives \textcolor{black}{are used}. The above equation is nothing but Eq.~(\ref{fesbia2}) and we therefore conclude that the set of Eqs.~(\ref{fesbia0})-(\ref{fesbia5}) are not independent. Hence the set of differential equations with which we shall deal consists of the two equations given in Eqs.~(\ref{eqsfmass}),(\ref{fesbia3}),(\ref{fesbia5}) and (\ref{massdef}). In order to find analytical solutions for this system, we firstly determine the equation of states in \textcolor{black}{the} radial and tangential directions. We assume \textcolor{black}{that} the fluid components obey linear equations of state, i.e., $p_r=w_r\rho$ and $p_\theta=w_\theta\rho$. Thus, the non-vanishing components of {the} effective EMT read
\bea
\rho^{\rm eff}&=&\f{(w_r+2w_\theta+3)\gamma-1}{4\gamma-1}\rho,\label{EFFEMTw1}\\
p_r^{\rm eff}&=&\f{(3w_r-2w_\theta+1)\gamma-w_r}{4\gamma-1}\rho,\label{EFFEMTw2}\\
p_\theta^{\rm eff}&=&\f{(2w_\theta-w_r+1)\gamma-w_\theta}{4\gamma-1}\rho.\label{EFFEMTw3}
\eea
Next, we assume \textcolor{black}{that} effective pressure in the collapsing configuration is zero, that is, $p_r^{\rm eff}=0$. This can be achieved once we set the Rastall parameter as 
\be\label{Rasdust} 
\gamma=\f{w_r}{3w_r-2w_\theta+1}.
\ee
Thus, from the second part of Eq.~(\ref{eqsfmass}), we readily find $F(t,r)=F(r)$. From Eq.~(\ref{fesbia5}) we can solve for the metric function $\nu$ with the solution given as
\be\label{nusol}
\nu(t,r)=\ln R(t,r)^{\f{2(w_\theta-w_r)}{w_r+1}}+F_1(t).
\ee
Substituting the above result into Eq.~(\ref{fesbia3}) and solving for the metric function $\psi$ we get
\be\label{psisol}
\psi(t,r)=\ln \left[R^\prime(t,r)R(t,r)^{\f{2(w_r-w_\theta)}{w_r+1}}\right]+F_2(r),
\ee
where $F_1(t)$ and $F_2(r)$ are arbitrary functions. It should be noted here that, though the effective pressure is zero and $F(t,r)=F(r)$, the solution is not necessarily same as the Lemaitre-Tolman-Bondi (LTB) metric~\cite{Joshibook}\cite{Plebanski}. However, the LTB spacetime can be recovered only for the case in which $w_r=w_\theta$ for which the metric (\ref{met}) reads
\be\label{psisolltb}
ds^2=-d\tilde{t}^2+\f{R'^2(\tilde{t},r)}{1+2E(r)}dr^2+R^2(\tilde{t},r)d\Omega^2,
\ee
where we have performed the transformation $d\tilde{t}=e^{F_1(t)}dt$ and $E(r)=(e^{-2F_2(r)}-1)/2$ is a free function so that the quantity $c^2E(r)/G$ plays the role of total energy within the same shell of matter~\cite{Plebanski}. The remaining equations i.e., first part of Eqs.~(\ref{eqsfmass}) and (\ref{massdef}) deal with the matter distribution and collapse dynamics, respectively. Let us now rescale the area radius function $R(t,r)$ so that at \textcolor{black}{the} initial time $t=0$ at which the collapse begins we have $R(0,r)=r$. Therefore, the first part of Eq.~(\ref{eqsfmass}) leaves us with the following integral for the mass function
\be\label{Frint}
F(r)=\kappa_{r}\int_{0}^{r}x^2\rho^{\rm eff}(0,x)dx.
\ee
Once the initial density of the collapsing body is specified, the mass function is determined. Finally, from Eq.~(\ref{massdef}) and solutions (\ref{nusol}) and (\ref{psisol}) we get the following equation for the dynamics of the area radius
\be\label{Reqdyn}
F(r)-R(t,r)-R(t,r)^\beta\dot{R}(t,r)^2f_1(t)+R(t,r)^\delta f_2(r)=0,
\ee
where
\bea\label{betadelta}
\beta&=&\f{5w_r-4w_\theta+1}{w_r+1},~~~~~\delta=\f{4w_\theta-3w_r+1}{w_r+1},\nn
F_1(t)&=&-\f{1}{2}\ln f_1(t),~~~~~F_2(r)=-\f{1}{2}\ln f_2(r).
\eea
\section{Nonsingular Collapse Solutions}\label{sec2}
Our aim in the present section is to build and study inhomogeneous collapse solutions for which the effective EMT in radial direction assumes a dust-like fluid but, due to non-minimal coupling between matter and geometry, the \textcolor{black}{the} matter content itself has nonzero pressure. In this regard, a class of nonsingular collapse solutions can be obtained by taking the relation $w_r=4w_\theta/5-1/5$ between \textcolor{black}{the} EoS parameters. Consequently Eq.~(\ref{Reqdyn}) reads (we choose the geometrized system of units so that $G=c=1$)
\be\label{eqR45}
f_1(t)\dot{R}^2-f_2(r)R^2+R-F(r)=0.
\ee
The above equation admits a general solution given by
\bea\label{gensol45}
R(t,r)\!\!&=&\!\!F(r){\rm exp}\left[f_2(r)^{\f{1}{2}}\left(\int_{1}^{t}\f{\pm dy}{\sqrt{f_1(y)}}+{\rm C}_1(r)\right)\right]-\f{1}{f_2(r)}\sinh\left(\f{f_2(r)^{\f{1}{2}}}{2}\left[\int_{1}^{t}\f{\pm dy}{\sqrt{f_1(y)}}+{\rm C}_1(r)\right]\right)^{\!\!2},\nn
\eea
where ${\rm C}_1(r)$ is an arbitrary function of radial coordinate $r$. In order to find this function we utilize the condition on rescaling the area radius, i.e., $R(0,r)=r$. This gives ${\rm C}_1(r)$ for each sign of Eq.~(\ref{gensol45}) as
\bea\label{C1expression}
{\rm C}_1(r)\!\!\!&=&\!\!\!\f{4\pi i{\rm C}_2}{\sqrt{f_2(r)}}\mp\!\!\int_{0}^{1}\!\!\f{dy}{\sqrt{f_1(y)}}+2\ln\left[-V(r)^{-\f{1}{2}}\right],\nn
{\rm C}_1(r)\!\!\!&=&\!\!\!\f{4\pi i{\rm C}_2}{\sqrt{f_2(r)}}\mp\!\!\int_{0}^{1}\!\!\f{dy}{\sqrt{f_1(y)}}+\ln\left[V(r)^{-\f{1}{2}}\right],
\eea
where ${\rm C}_2\in{\mathbb Z}$ and
\bea\label{V}
\!\!\!\!V(r)=1\pm2\sqrt{f_2(r)\left[F(r)+r^2f_2(r)-r\right]}-2rf_2(r).
\eea
Substituting the arbitrary functions ${\rm C}_1(r)$ \textcolor{black}{from (\ref{C1expression})} into Eq.~(\ref{gensol45}) we arrive at eight expressions as solutions to equation (\ref{eqR45}) among which only two of them are independent. Hence, the final solution is obtained as (we have set ${\rm C}_2=0$)
\bea\label{solclass1}
R(t,r)\!\!&=&\!\!\f{F(r)}{V(r)}{\rm exp}\left(-f_2(r)^{\f{1}{2}}\int_{0}^{t}\f{dy}{\sqrt{f_1(y)}}\right)-\f{1}{f_2(r)}\sinh\left(\f{f_2(r)}{2}^{\f{1}{2}}\!\!\!\!\int_{0}^{t}\f{dy}{\sqrt{f_1(y)}}+\ln\left[-V(r)^{\f{1}{2}}\right]\right)^2.\nn
\eea
Next we proceed to find the mass function by setting a suitable form for the initial energy density. To this aim we use the first part of Eq.~(\ref{eqsfmass}) along with Eqs.~(\ref{EFFEMTw1}) and (\ref{Rasdust}) at $t=0$. We therefore get the following differential equation for the mass function
\be\label{diffeqmass}
\f{32\pi r(w_\theta+1)(7w_\theta-3)\epsilon(r)}{11w_\theta-4}-\f{5F^\prime(r)}{r}=0,
\ee
where $\rho(t=0,r)=\epsilon(r)$ being the initial profile of the energy density for which, we take the following {\it ansatz}\footnote{This form of initial energy density has been inspired by the works~\cite{inhommods11}\cite{iniprofen}\cite{Goni2023}, however, other types of initial profiles for energy density have been suggested in the literature~\cite{otherdenprofs}. One may also find similarities between the density profile (\ref{epsils}) and those proposed for superdense stars, see e.g.~\cite{superdensprofs}.}
\bea\label{epsils}
\epsilon(r)\!\!&=&\!\!\rho_0\left[\f{2r_b^n}{r_b^n+r^n}-1\right],
\eea
where $\rho_0$ is the energy density at the center $(r=0)$ of the collapsing object. The mass function then reads
\bea\label{massfunc}
&&F(r)=\f{{64\pi} r^3\rho_0(w_\theta+1)(7w_\theta-3)}{5(33w_\theta-12)}\left({_2}F_1\left[1,\f{3}{n},\f{n+3}{n},-\f{r^n}{r_b^n}\right]-\f{1}{2}\right).
\eea
\begin{figure}
	\begin{center}
		\includegraphics[width=7.7cm]{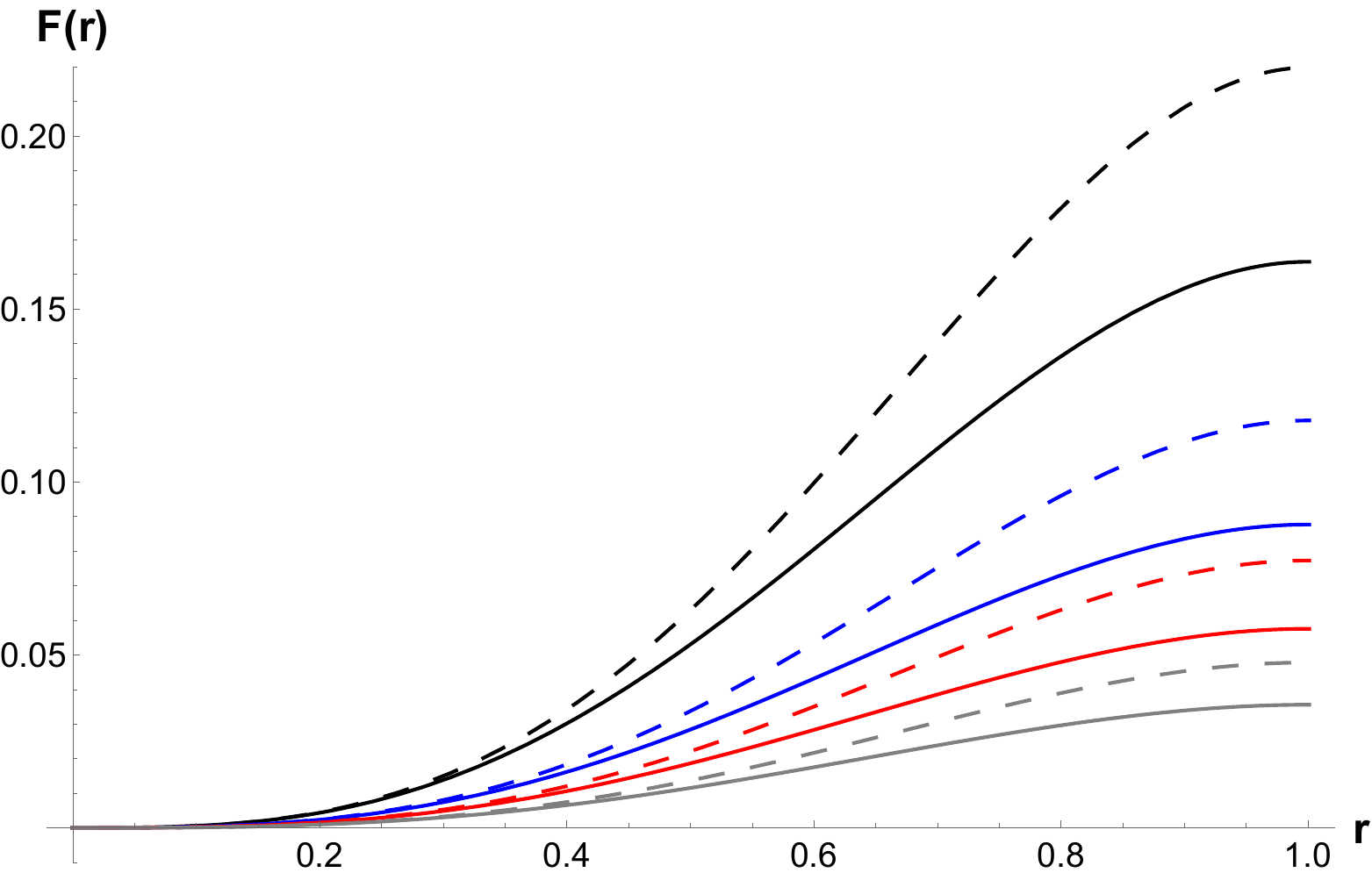}
		\caption{Plot of mass function for $8\pi\rho_0=1$, $r_b=1$, $n=2$ (family of solid curves) and $n=3$ (family of dashed curves). The EoS in tangential direction has been set as $w_\theta=1/3,1,0,-1/3$ corresponding to black, blue, red and gray curves.}\label{fig1}
	\end{center}
\end{figure}
The mass function shows a monotonically increasing function of radial coordinate which vanishes at the center, see Fig.~(\ref{fig1}). A point that is worth mentioning here is that the variables in geometrized units are shown as powers of length [${\sf m}$]~\cite{MTWbook}. For SI units of $[{\sf kg}^j{\sf m}^\ell{\sf s}^q]$, the conversion factor from geometrized units of ${\sf m}^{j+\ell+q}$ to SI units is given by, $G^{-j}c^{2j-q}$~\cite{schutzbook}. Hence, the initial value of energy density at the center in geometrized units is equal to the following value in SI units
\bea\label{rho0geosi}
\rho_0=\f{1}{8\pi}\left[{\sf m}^{-2}\right]\left(G^{-1}c^4\right)=\f{\left(2.9979\times10^8{\sf m}/{\sf s}\right)^4}{\left(6.6743\times10^{-11}{\sf m}^5/{\sf kg}{\sf s}^2\right)}=0.048\times10^{44}\left[\f{{\sf kg}}{{\sf m}{\sf s}^2}\right].
\eea
To find the mass density at the center we need to divide the above result by the square of the speed of light. A straightforward calculation then gives
\be
\rho_0/c^2=5.3\times10^{25}\left[\f{{\sf kg}}{{\sf m}^3}\right]=5.3\times10^{22}\left[\f{{\sf gr}}{{\sf cm}^3}\right].
\ee
The above mass density can be compared to that of {\it Preon stars}, hypothetical compact objects that could have been formed by primordial density fluctuations in the early universe, and in the \textcolor{black}{collapse} process of massive stars~\cite{Preonst}\cite{Preonst1}\cite{Preonst2}\cite{CPDM1}. Moreover, the conversion factor for SI units of $\left[{\sf kg}\right]$ is $c^2G^{-1}$, then, for an object with unit mass in geometrized units we get
\bea\label{massgusi}
1[{\sf m}]c^2G^{-1}=\f{\left(2.9979\times10^8{\sf m}/{\sf s}\right)^2}{\left(6.6743\times10^{-11}{\sf m}^3/{\sf kg}{\sf s}^2\right)}=1.3466\times10^{27}\left[{\sf kg}\right],
\eea
whence with the help of Eq.~(\ref{massfunc}) one can estimate the total mass of the compact object in comparison to the solar mass and the Earth mass, as~\cite{Preonst}\cite{Preonst2}
\bea\label{mass23}
\!\!\!\!\!\!\!\!\!&M&\!\!\!\!_{\rm tot}\!\simeq(0.2\!-\!1.3)\!\times\!10^{-4}M_\odot\!\simeq\!(8\!-\!46)M_\oplus,~n=2,\\
\!\!\!\!\!\!\!\!\!&M&\!\!\!\!_{\rm tot}\!\simeq(0.3\!-\!1.8)\!\times\!10^{-4}M_\odot\!\simeq\!(10\!-\!61)M_\oplus,~n=3.
\eea
We further note that the conversion factor for the SI unit of length is unity. To find the above results we have taken the initial radius of the compact object as $R(0,r_b)=r_b=1$ $[{\sf m}]$ in geometrized units which is equal to unit length in SI units. This is in accordance with the maximum radius of a star composed of Preons\footnote{Hypothetical sub-quark particles proposed by some Beyond the Standard Model theories in which, quarks, leptons, and sometimes some of the gauge bosons, are composite particles built out of such  particles~\cite{Preonspart}\cite{Preonsparta}\cite{Preonspartb}}, i.e., $R_{\rm max}\sim1$, which has been previously reported in~\cite{Preonst}\cite{Preonst2}. Therefore, in the present model, the stellar system that undergoes the collapse process can be considered as a Preon star with an ultra-dense core. These hypothetical compact objects are classified as exotic stars, which are not composed of the standard matter found in \textcolor{black}{the} white dwarfs and neutron stars~\cite{Preonst2}. The possible mechanism that could trigger the collapse of such a compact object may be due to the failure of Preon degeneracy pressure. In this case, if fermionic Preons exists~\cite{Preonspart}, \textcolor{black}{the} star mass could exceed a Preon-equivalent Tolman-Oppenheimer-Volkoff limit, due to accretion of additional mass from a companion or merger with another Preon star. Consequently, if the Preon degeneracy pressure fails to stabilize the star against the extreme pull of gravity, it may undergo a continual collapse process. Also, in \textcolor{black}{the} cosmological contexts, strange matter distribution called, {\it compact Preon dark matter} has been proposed as a possible dark matter candidate. These exotic objects may have been created in a first-order phase transition in the early universe~\cite{Preonst2}\cite{CPDM1}\cite{CPDM}\cite{Ponomarev2007}\cite{Vereshkov1999}.%that the core of the star exerts on its outer layers. Ponomarev possibility
\par 
Let us now consider the following forms for the free functions
\bea\label{freeff}
f_1(y)=\alpha,~~~~~~~~f_2(r)=\f{\xi}{r+\zeta},
\eea
where $(\alpha,\xi)\in{\mathbb R}$ are constants and $0<\zeta\ll1$ has been chosen so that the spacetime metric be regular at $r=0$. Utilizing then expression~(\ref{massfunc}) for the mass function, we are able to evaluate the behavior of the area radius against time for each collapsing shell. By doing so, we get the following expression for the area radius
\bea\label{arearad}
R(t,r)&=&\f{32\pi\rho_0r^3(1+w_\theta)(7w_\theta-3){\rm e}^{\left[-t\sqrt{\f{\xi}{\alpha(r+\zeta)}}\right]}}{5(33w_\theta-12)g(r)}\left({_2}F_1\left[1,\f{3}{n},\f{n+3}{n},-\f{r^n}{r_b^n}\right]-\f{1}{2}\right)\nn&-&\f{r+\zeta}{\xi}\sinh\left[\f{t}{2}\sqrt{\f{\xi}{\alpha(r+\zeta)}}-\ln\left[\f{-1}{\sqrt{g(r)}}\right]\right]^2,
\eea
where
\bea\label{gfunct}
\!\!\!\!g(r)=1-\f{2\xi r}{r+\zeta}+2\Bigg(\f{\xi r}{r+\zeta}\Bigg[-1+\f{\xi r}{r+\zeta}+\f{32\pi\rho_0r^2(1+w_\theta)(7w_\theta-3)}{5(33w_\theta-12)}{_2}F_1\left[1,\f{3}{n},\f{n+3}{n},-\f{r^n}{r_b^n}\right]-\f{1}{2}\Bigg]\Bigg)^{\f{1}{2}}.\nn
\eea
\begin{figure}
	\begin{center}
		\includegraphics[width=7.7cm]{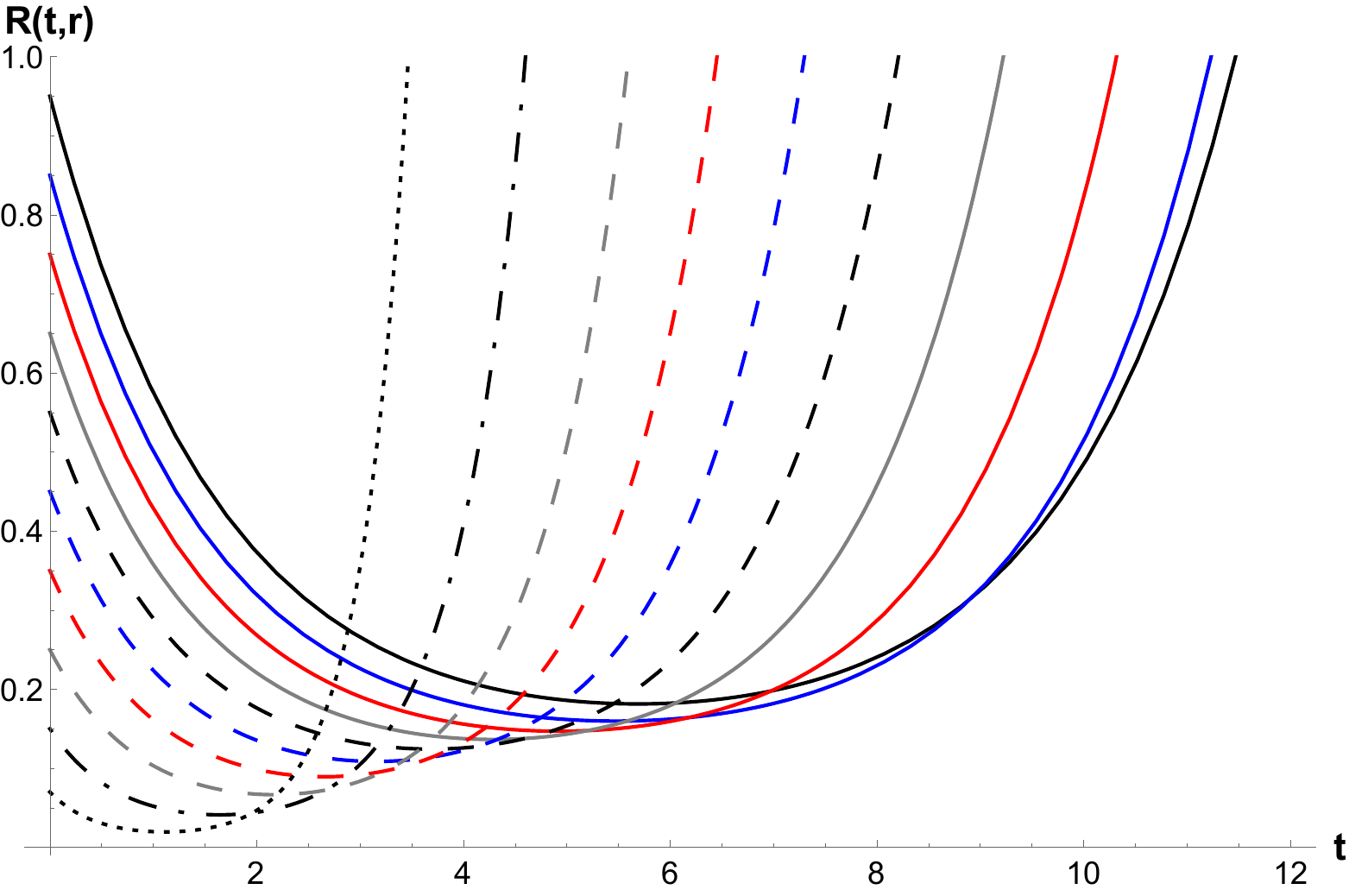}
		\caption{Evolution of the area radius for $8\pi\rho_0=1$, $r_b=1$, $n=2$, $\zeta=0.001$, $\alpha=10$, $\xi=3.6$ and $w_\theta=0$. The matter shells are indicated by different values of radial coordinate so that, the solid curves with black, blue, red and gray colors correspond to $r=\{0.95,0.85,0.75,0.65\}$ respectively. Those for which $r=\{0.55,0.45,0.35,0.25\}$ correspond to dashed curves with the very order of the colors and, $r=\{0.15,0.05\}$ correspond to dot-dashed and dotted curves.}\label{fig2}
	\end{center}
\end{figure}
\begin{figure}
	\begin{center}
		\includegraphics[width=7.7cm]{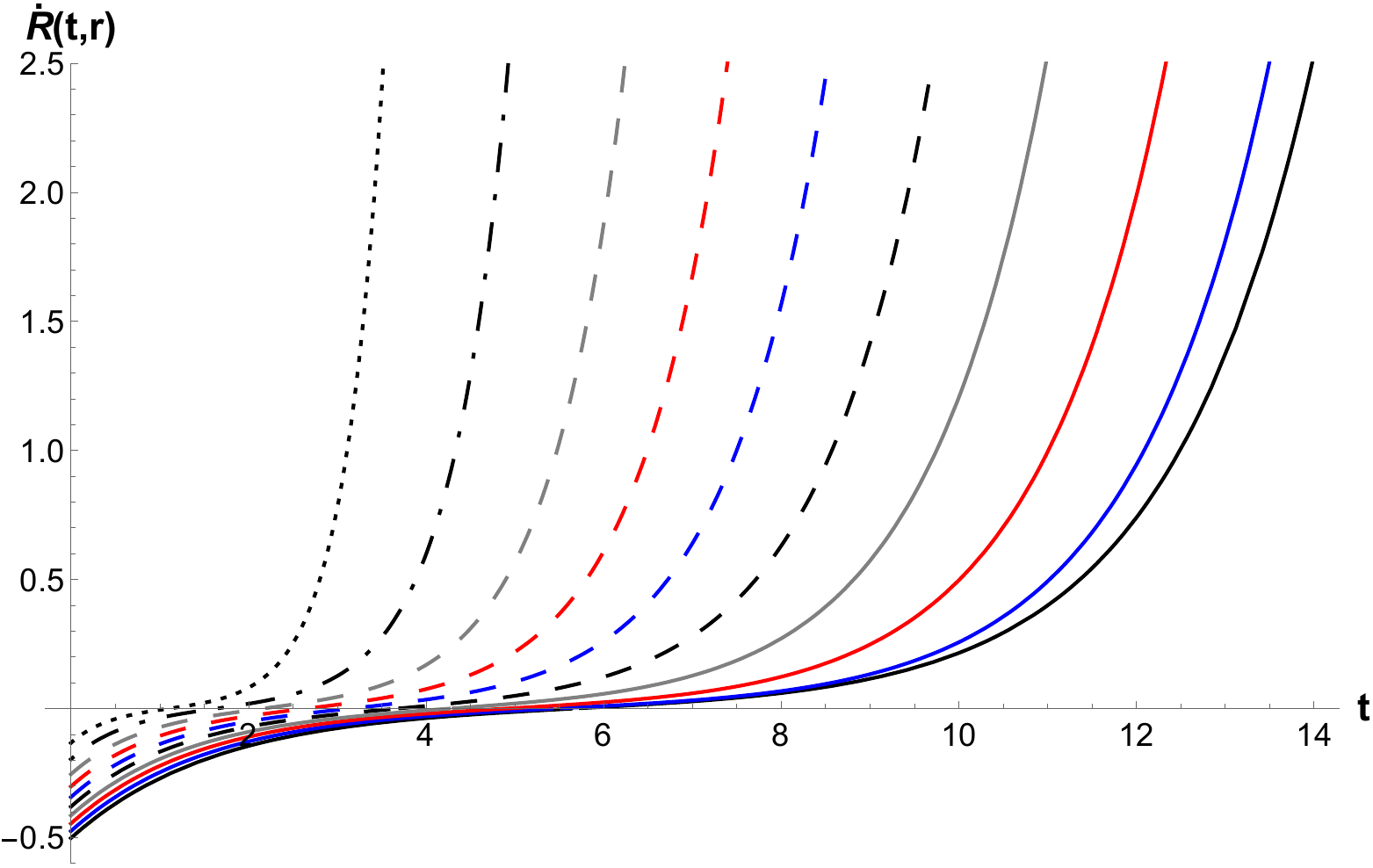}
		\caption{Evolution of collapse velocity for the same values of model parameters as of Fig.(\ref{fig2}).}\label{fig3}
	\end{center}
\end{figure}
Figure~(\ref{fig2}) shows the behavior of area radius against time for different values of the shell radii. We have considered the case with $n=2$ and a dust fluid in tangential direction, i.e., $w_\theta=0$. We therefore observe that for each collapsing shell, the area radius commences its evolution according to the condition $R(0,r)=r$ and after a while reaches a minimum value at the bounce time, $t_{\rm b}(r)$, where $\dot{R}(t_{\rm b}(r),r)=0$. After the bounce, each shell enters an expanding phase and disperses away without a singular end-state. In Fig.~(\ref{fig3}) we have sketched the evolution of collapse velocity for each shell which further verifies such a behavior. It is seen that a shell of matter starts its evolution in a contracting regime where $\dot{R}(t_{\rm b}(r),r)<0$ for $t<t_{\rm b}(r)$. \textcolor{black}{Next}, the collapse halts at the bounce point, and an expanding regime begins its evolution where $\dot{R}(t_{\rm b}(r),r)>0$ for $t>t_{\rm b}(r)$. The closer the shells to the center, the sooner they bounce back to an expanding phase. Figure (\ref{fig4}) shows the behavior of infinitesimal difference in area radii between two shells labeled by $r$ and $r+dr$ at a given instant of time. We first note that $R^\prime(t_\star(r),r)=0$ where $t_\star(r)$ is the time at which two infinitesimal shells intersect, hereafter we call it the {\it shell intersection time}. For $t<t_\star(r)$ and those layers whose characteristic curves are shown in blue to black dotted ones ($r=0.85\rightarrow0.05$), the infinitesimal difference in area radius is a positive decreasing function over time and for $t>t_\star(r)$, it is a decreasing \textcolor{black}{one} in negative direction. This means that, as the collapse proceeds, the difference between matter shells decrease until the shell intersection time is reached, after which, the shells cross each other so that the inner shells supersede the outer ones and begin to move away. This is why $R^\prime(t,r)<0$ and $R^\prime(t,r)\rightarrow-\infty$ (the matter shells disperse to infinity) for $t>t_\star(r)$. However, there exist outer shells~\footnote{The specified numerical range for shell coordinate radii corresponds to the selected model parameters for which Fig.~(\ref{fig2}) has been plotted. Other values for model parameters may give different ranges.} ($0.933\lessapprox r\leq1$) for which $R^\prime(t,r)$ reaches a minimum value and then admits a maximum before getting vanished at the shell crossing time. These outer layers behave differently with respect to previous ones and their behavior can be summarized in four phases: $i)$ As the collapse commences, the shells begin to approach each other until reaching a minimum difference, $ii)$ the shells move a way from each other to a finite maximum distance, $iii)$ they then start approaching till they meet each other at the shell crossing time, $iv)$ after the inner shells get replaced by the outer ones, they finally move away from each other and disperse. The situation that occurs at $t_\star(r)$ is referred to as the occurrence of a {\it shell-crossing singularity}, where, nearby shells of matter intersect creating thus temporary density singularities along with divergences in some of the curvature scalars~\cite{shellcr}. The shell-crossing singularities are generally known as weak in curvature strength~\cite{clanol} in contrast to \textcolor{black}{the} strong curvature {\it shell-focusing singularities} such as the Big-Bang and the curvature singularity at the center of the Schwarzschild spacetime, see e.g.,\cite{Joshibook}\cite{COUNTERCCC}\cite{Plebanski} and discussions therein. A question that may arise here is: Do the shell-crossings occur after or before the bounce time? To answer this question we proceed with evaluating the bounce time which can be expressed as the root of the equation $\dot{R}(t_{\rm b}(r),r)=0$. Considering the model parameters for which Fig.~(\ref{fig2}) has been plotted we find that there exist four roots for this equation that the two of which gives negative values and the third one gives complex values for the bounce time. The only physically reasonable root is then given by 
\bea\label{bounct}
t_{\rm b}(r)=\sqrt{\f{\alpha(r+\zeta)}{\xi}}\cosh^{-1}\!\!\left[\f{\sqrt{5}(r-2\xi r+\zeta)}{\sqrt{(r+\zeta)h(r)}}\right],
\eea
where
\be\label{hfunc}
h(r)=r\left[4\xi(r^2-6)+5\right]+24\xi\tan^{-1}(r)+5\zeta.
\ee
To find the shell intersection time, we have to solve the equation $R^\prime(t_\star(r),r)=0$, however, as this equation is not algebraic we cannot solve it using standard methods. We therefore resort to numerical techniques to find the shell intersection time as a function of radial coordinate. The result is shown in Fig.~(\ref{fig5}), where we observe that for all matter shells $t_\star(r)>t_{\rm b}(r)$. This means that the collapsing shells experience the bounce event before they cross each other, or equivalently the shell-crossing singularity occurs in the expanding post-bounce regime where, the inner shells surpass the outer ones. 
\par
Finally, to deal with the WEC, we consider the first part of Eq.~(\ref{eqsfmass}) together with Eqs.~(\ref{EFFEMTw1}) and (\ref{Rasdust}) which lead to the following expression at the onset of collapse
\be\label{rhoeffonset}
\rho^{\rm eff}(0,r)=\f{4}{5}\rho(0,r)=\f{9\rho_0(r_b^n-r^n)}{20\kappa(r_b^n+r^n)},~~~~w_\theta=0,
\ee 
whereby we figure out that, the effective energy density and the fluid energy density are positive provided that $\rho_0>0$, as required by the regularity of initial conditions~\cite{Joshibook}. As the collapse progresses, the energy density of the fluid and the effective one, grow until reaching the bounce point where both densities assume maximum positive values. We note that since $t_{\rm b}(r)<t_\star(r)$, then $\rho^{\rm eff}(t,r)>0$ and $\rho(t,r)>0$ as long as $t<t_\star(r)$. At the shell crossing time the densities diverge (since $R^\prime(t_\star(r),r)=0, F^\prime\neq0$) and they become negative for $t>t_\star(r)$. In other words, in the post bounce regime, the fluid and effective energy densities admit negative values because the less dense layers of outer regions take the place of more dense inner layers. Therefore, the positivity of energy density is fulfilled in the entire contracting phase and during the finite time interval $\Delta t(r)=t_\star(r)-t_{\rm b}(r)$ in the expanding phase. Regarding the conditions (\ref{WEC}) and (\ref{WECeff}), we find that same discussions can be made for radial and tangential profiles of the WEC as
\bea\label{discwec}
\rho^{\rm eff}\!\!+p^{\rm eff}_r=\rho+p_r=\f{4}{5}\rho,~~~\rho^{\rm eff}\!\!+p^{\rm eff}_\theta=\rho+p_\theta=\rho.
\eea
Also for the DEC, the second part of Eq.~(\ref{decsec1}) gives $\rho\geq|w_r\rho|$ and $\rho\geq|w_\theta\rho|$ whence, for the current choice of EoS parameters we get $\rho\geq|-\rho/5|$ and $\rho\geq0$. We therefore find that for $t<t_\star(r)$ the first and second parts of Eq.~(\ref{decsec1}) hold and hence the DEC is satisfied while it is violated for $t>t_\star(r)$. The same discussion also follows for the effective form of DEC given in Eq.~(\ref{decseceff1}). Finally, for the SEC the second part of Eqs.~(\ref{decsec2}) and (\ref{decseceff2}) give
\bea\label{secpresent}
&&\!\!\!\!\!\!\!\!\!\rho+\sum_{i}p_i\geq0=\rho+p_r+2p_\theta=\f{2}{5}\left(2+7w_\theta\right)\rho=\f{4}{5}\rho,\nn
&&\!\!\!\!\!\!\!\!\!\rho^{\rm eff}+\sum_{i}p_i^{\rm eff}\geq0=\rho^{\rm eff}+p_r^{\rm eff}+2p_\theta^{\rm eff}=\f{6}{5}\left(1+w_\theta\right)\rho=\f{6}{5}\rho.\nn
\eea
It is therefore seen that the SEC is satisfied for $t<t_\star(r)$ and gets violated for $t>t_\star(r)$.
\begin{figure}
	\begin{center}
		\includegraphics[width=7.7cm]{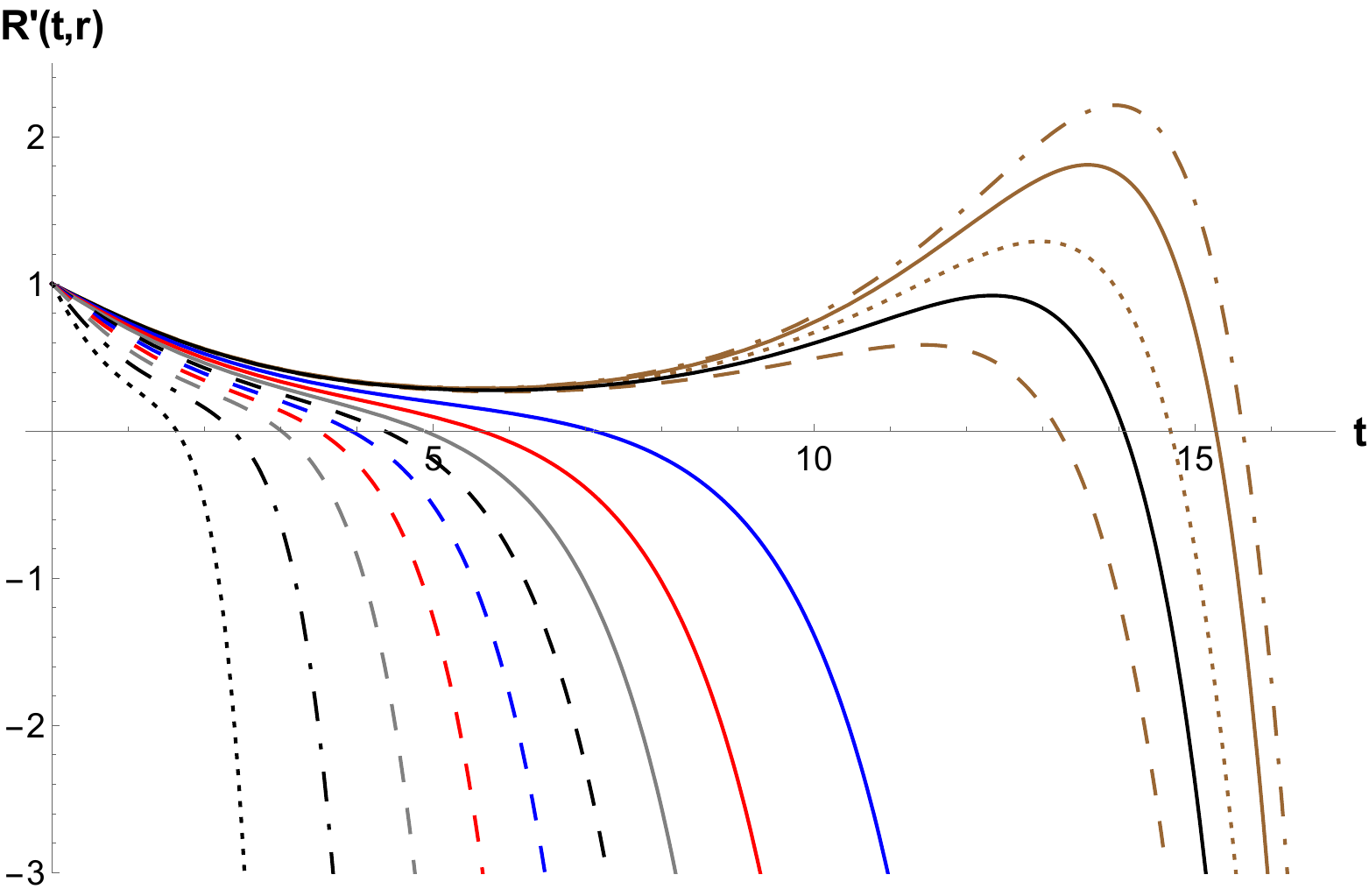}
		\caption{Evolution of the difference between physical radii of the shells for the same values of model parameters as of Fig.(\ref{fig2}). The extra four brown curves have been plotted assuming the coordinate radii $r=\{0.963,0.955,0.943,0.96\}$ for dot-dashed, dotted, dashed, solid curves, respectively.}\label{fig4}
	\end{center}
\end{figure}
\begin{figure}
	\begin{center}
		\includegraphics[width=7.7cm]{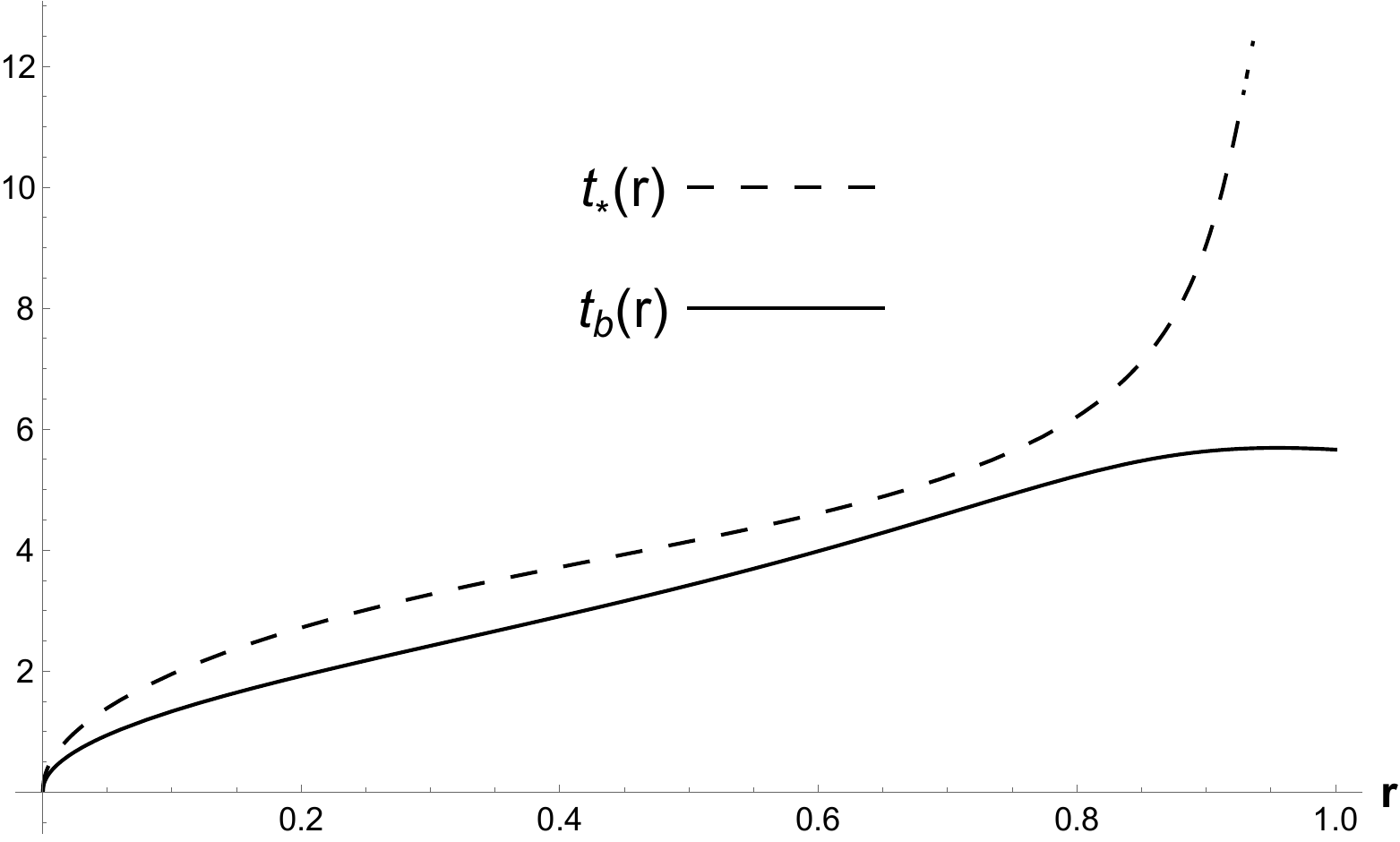}
		\caption{Behavior of the bounce and shell crossing times against radial coordinate.}\label{fig5}
	\end{center}
\end{figure}
\section{Trapped Surfaces}\label{sec3}
\begin{figure}
	\begin{center}
		\includegraphics[width=7.7cm]{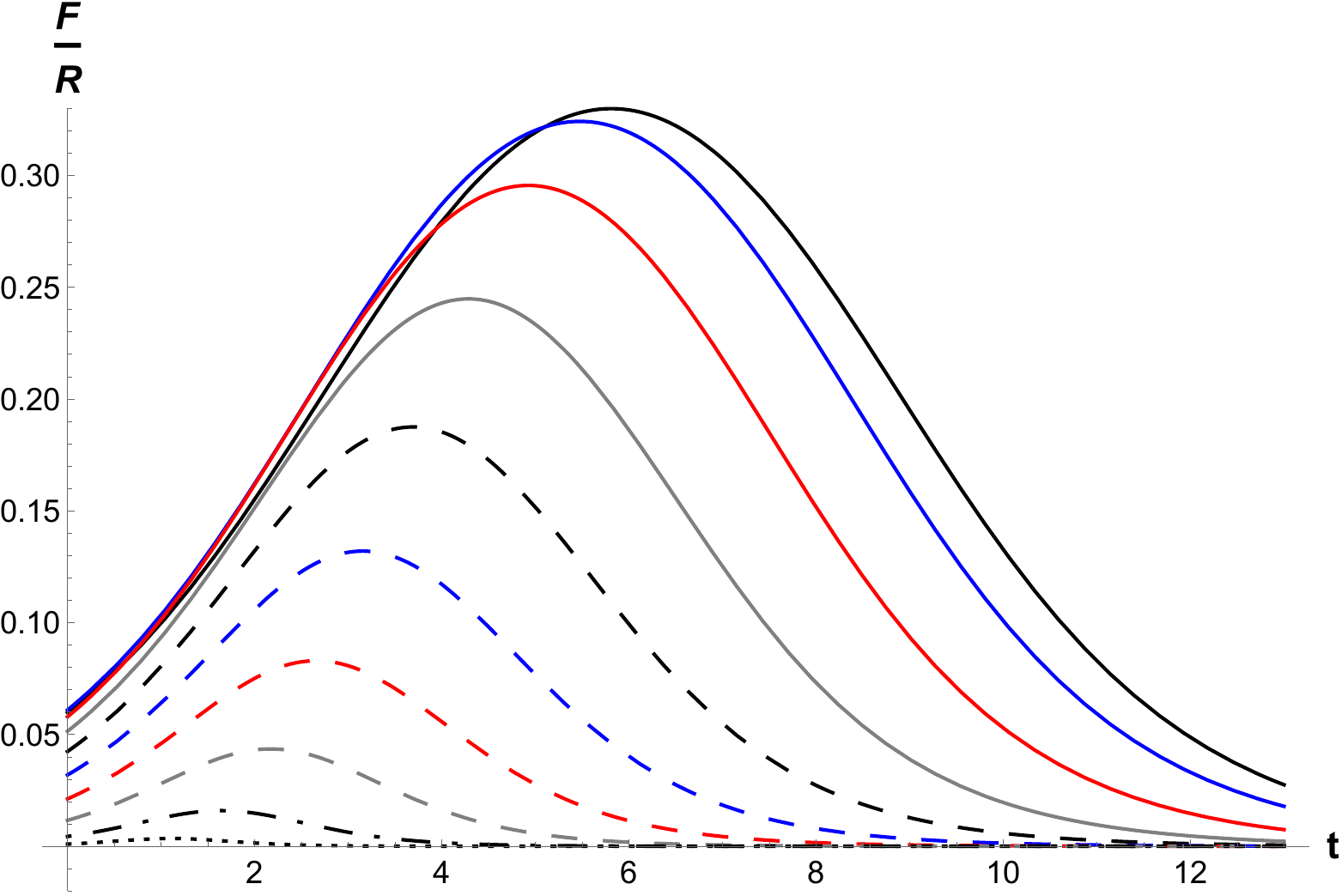}
		\caption{Behavior of the ratio $F/R$ during the entire evolution of the collapsing object.}\label{fig6}
	\end{center}
\end{figure}

An important issue that deserve more investigation is to analyze the formation of the trapped surfaces and apparent horizon in the collapse scenario. Indeed, such analysis could specially help to recognize whether the bounce is visible or not. Trapped surfaces are defined as compact two-dimensional space-like surfaces in such a way that both families of ingoing and outgoing null geodesics normal to them necessarily converge~\cite{Joshibook}\cite{trapsurf}. To deal with \textcolor{black}{the} trapped surface formation for the present collapse scenario, we follow the terminology presented in~\cite{MSEE}. Using then the metric potentials Eqs.~(\ref{nusol}) and (\ref{psisol}), the line element (\ref{met}) can be rewritten as
\be\label{met1}
ds^2=-\f{R}{\alpha}dt^2+\f{R^{\prime2}(r+\zeta)}{R\xi}dr^2+R^2d\Omega^2.
\ee
If we introduce the null coordinates
\bea\label{nullcoords}
d\eta^{\pm}&=&-\frac{1}{\sqrt{2}}\left[\sqrt{\f{R}{\alpha}}dt\mp\sqrt{\f{r+\zeta}{R\xi}}R^\prime dr\right],
\eea
the line element (\ref{met1}) can be recast into double-null form as
\be\label{metricdnull}
ds^2=-2d\eta^{+}d\eta^{-}+R(t,r)^2d\Omega^2.
\ee
\textcolor{black}{the} radial null geodesics in the spacetime with the above line element are given by the condition $ds^2=0$, whereby, we recognize that there exist two kinds of future-directed null geodesics which correspond to $\eta^{+}=constant$ and $\eta^{-}=constant$. The expansion parameters along these geodesics are defined as
\be\label{expansion}
\Theta_{\pm}=\frac{2}{R(t,r)}\frac{\partial}{\partial\eta^{\pm}}R(t,r),
\ee
where the differential operators in null direction are given by
\bea\label{nulldiffs}
\f{\partial}{\partial\eta^{\pm}}=\f{1}{\sqrt{2}}\left[\sqrt{\f{\alpha}{R}}\partial_{t}\mp\sqrt{\f{R\xi}{r+\zeta}}\f{\partial_r}{R^\prime}\right].
\eea
We therefore get the expansion parameters as
\bea\label{expparms}
\Theta_{\pm}=\f{\sqrt{2}}{R}\left[\sqrt{\f{\alpha}{R}}\dot{R}\mp\sqrt{\f{R\xi}{r+\zeta}}\right].
\eea
The expansion parameters measure whether a family of null geodesics normal to a sphere is diverging $(\Theta_{\pm}>0)$ or converging $(\Theta_{\pm}<0)$, or equivalently, the area radius along such geodesics is increasing or decreasing, respectively. A spacetime is referred to
as trapped, untrapped and marginally trapped if
\be\label{sptruntr}
\Theta_{+}\Theta_{-}>0,~~~~~~ \Theta_{+}\Theta_{-}<0,~~~~~~\Theta_{+}\Theta_{-}=0,
\ee
respectively, where the third class implies the outermost boundary of the trapped region, i.e., the apparent horizon. Using equations (\ref{eqR45}) and (\ref{freeff}) we readily find that the quantity $\Theta_{+}\Theta_{-}$ is related to mass function as
\be\label{thetamassf}
\Theta_{+}\Theta_{-}=\frac{2}{R^2(t,r)}\left[\frac{F(r)}{R(t,r)}-1\right].
\ee
Therefore, \textcolor{black}{the} conditions (\ref{sptruntr}) can be rewritten as
\be\label{condstrappedness}
\f{F(r)}{R(t,r)}>1,~~~~~\f{F(r)}{R(t,r)}<1,~~~~~\f{F(r)}{R(t,r)}=1,
\ee
where the equality provides us with the location of apparent horizon. In view of the above conditions, we realize that if at the onset of the collapse, the ratio $F(r)/R(t_i,r)<1$ (as required by the regularity of initial conditions), and stays less than unity, then the apparent horizon is failed to form during the dynamical evolution of the collapsing body. If however, the ratio $F(r)/R(t,r)$ becomes greater than unity, then the apparent horizon is formed to cover the bounce. Figure~(\ref{fig6}) shows the behavior of $F/R$ against time where it is seen that this ratio is less than unity at $t=t_i$ when the collapse commences. Hence, the condition on absence of trapped surfaces at initial epoch is satisfied. We also observe that $\Theta_{+}\Theta_{-}<0$ for $t>t_i$ which implies that \textcolor{black}{the} trapped surfaces are failed to form in the entire dynamical evolution of the collapsing body and consequently, the bounce event is not covered by the apparent horizon.
\section{concluding remarks}\label{sec4}
Despite its successes~\cite{GRSUC}, GR, like other physical theories is not the ultimate theory that describes gravitational interactions, as it suffers from the occurrence of spacetime singularities. The true curvature singularities of GR give rise to \textcolor{black}{the} fundamental challenges, as they represent spacetime events where our current understanding of the Universe fails, leading to what many physicists deem an {\it unpleasant situation} in theoretical physics~\cite{Hawpenro}. Much efforts to address the problems caused by singularities have been made so far at both classical and quantum levels. \textcolor{black}{The} classical attempts often involve modified gravity theories that extend or alter the predictions of GR to avoid singularities~\cite{FRsingbamba}\cite{modgrnonsing}\cite{nonsinmodif}. At \textcolor{black}{the} quantum levels, theories such as loop quantum gravity and string theory suggest mechanisms that could potentially resolve singularities by introducing new dimensions or quantized spacetime structures see e.g.,~\cite{singremove}\cite{quantumcollapse}\cite{LQCSTRING}. These theories aim to provide a more complete understanding of gravitational interaction, particularly in extreme environments like the very final stages  of a collapse scenario, where the known laws of physics fail.
\par
In the present work, we tried to build and study inhomogeneous collapse models in \textcolor{black}{the} Rastall gravity for which the spacetime singularity (present in GR models) is replaced by a nonsingular bounce. In this sense, the matter shells that constitute the collapsing object, reach a minimum physical radius at a finite amount of time and then rebound to infinity without ever forming a shell-focusing singularity. We therefore observed that the speed of collapse for each shell begins from a negative value in the contracting regime, vanishes at the bounce time and then assumes positive values at the post-bounce regime. We further noticed that a density singularity occurs at $R^\prime=0$, which is due to \textcolor{black}{the} shell-crossings. Basically, these type of singularities indicate the breakdown of the coordinate system used and are not generally regarded as genuine spacetime singularities~\cite{YodMuller1973}, because they can be possibly removed from the spacetime by extending the manifold~\cite{clanol}\cite{clarke1986}. By examining the bounce and shell intersection times we found that since $t_\star(r)>t_{\rm b}(r)$, the shell-crossing singularities appear in the post-bounce regime where \textcolor{black}{the matter} shells explode in such a way that \textcolor{black}{the} fast moving inner shells overtake the outer \textcolor{black}{ones} and then disperse to infinity. The violation of WEC in the post-bounce regime, where the effective energy density becomes negative, is a direct consequence of the shell-crossings. However, the negativity of the effective energy density arises from the breakdown of the comoving coordinate description at shell crossing, rather than an intrinsic pathology of the matter field~\cite{clanol}. The occurrence of such event, which has been studied in inhomogeneous collapse models, does not necessarily imply unphysical matter, but rather a limitation of the fluid description in the post-bounce regime~\cite{shellcross}. See also~\cite{loopcolshell} for the occurrence of shell-crossings in quantum corrected LTB collapse models. We finally examined trapped surface formation during the collapse process and concluded that the bounce is not covered by the apparent horizon. A point here that begs more consideration is that it is possible to examine other forms for the free functions Eq.~(\ref{freeff}) in order to avoid \textcolor{black}{the} shell-crossing singularities. If so, we may have the following considerations for the behavior of $R^\prime$ against time: $i)$ it remains positive throughout the dynamical evolution of the collapsing body, $ii)$ it \textcolor{black}{decreases} to a minimum value at or before the bounce where the matter shells reach the closest physical distance without intersection, $iii)$ and then it starts to increase to infinity. This means that, without overtaking each other, the matter shells move away with the speed determined by the acceleration of the post bounce process. We therefore conclude that in the context of \textcolor{black}{the} Rastall gravity, the spacetime singularity can be avoided in the collapse process of an inhomogeneous fluid. Despite many simplifications assumed in this work, it still may contribute to a better understanding of \textcolor{black}{the} gravitational collapse beyond \textcolor{black}{the} classical GR and may open up new avenues for probing the collapse \textcolor{black}{phenomenon} in \textcolor{black}{the context of} alternative theories of gravity. The Rastall parameter in the herein model plays a crucial role for generating exact nonsingular collapse solutions while the appropriate choice for the EoS parameters through Eq.~(\ref{Rasdust}) can provide consistent values to observational data for this parameter. Needless to say, much work is needed to connect the toy model investigated in this work, to \textcolor{black}{the} astrophysical observations such as, gravitational waves~\cite{grwaves} or black hole shadows~\cite{shadowbh}. Future research should explore observable \textcolor{black}{implications} by incorporating more realistic equations of state, radiation effects, and numerical simulations to \textcolor{black}{make the} Rastall gravity a theory with potential astrophysical applications. Finally, we would like to point out that the dynamical stability of the bounce solution under small perturbations represents a crucial aspect of its physical viability that requires careful investigation. While our current analysis has established the existence of the bounce solution within \textcolor{black}{the} Rastall framework, we acknowledge that a proper stability analysis through perturbations e.g., $\delta R(t,r)$ of the area radius or $\delta\rho(t,r)$ of the energy density particularly around the critical bounce epoch is indeed necessary to ensure its viability and robustness. A complete treatment of this issue is accompanied by technical and conceptual challenges such as, non-linear coupling of perturbations~\cite{grwaves}, gauge issues in inhomogeneous collapse~\cite{nolangauge}, formation of horizon and causal structure~\cite{batachairaban}, instability from negative energy density~\cite{insnegenden}, background dependence on shell-crossings~\cite{Joshibook}\cite{loopcolshell} and numerical errors near the bounce event~\cite{numerror}, see also~\cite{pertbounce} for more details. In this regard, the dynamical stability of the present solutions is an open question that begs a dedicated investigation. This will be the focus of future work, where we plan to employ both analytical methods and numerical techniques for simplified cases and a fully inhomogeneous problem, respectively.
\par
\vspace{0.5cm}
{\bf Data Availability Statement}: The data that support the findings of this study are available from the corresponding author, [AJ], upon reasonable request.

\end{document}